\documentclass[twocolumn,prl,showpacs,superscriptaddress]{revtex4}

\usepackage{graphicx}
\usepackage{dcolumn}
\usepackage{bm}

\begin{document}

\title{A Compressed Particle-Hole Symmetric Pfaffian State for $\nu = 5/2$ Quantum Hall Effect}
\author{Jian Yang}
\email{jyangmay1@yahoo.com}
\altaffiliation{Permanent address: 5431 Chesapeake Place, Sugar Land, TX 77479, USA}
%\affiliation{}
%\date{}

\begin{abstract}

A recent thermal Hall conductance experiment [Banerjee et al., Nature {\bf559}, 205 (2018)] for $\nu = 5/2$ fractional quantum Hall system appears to rule out both the Pfaffian and anti-Pfaffian and be in favor of the PH-Pfaffian topological order, while the existing numerical results without disorder have shown otherwise. In this paper we offer a possible resolution by proposing a new state, termed compressed PH-Pfaffian state by "compressing" the PH-Pfaffian state with two flux quanta removed to create two abelian Laughlin type quasiparticles of the maximum avoidance from one another (or of the maximum number of zeros). The compressed PH-Pfaffian state is not particle-hole symmetric but possesses the PH-Pfaffian topological order. In spherical geometry, the compressed PH-Pfaffian state has the same magnetic flux number $N_{\phi}= 2N-3$ as the Pfaffian state, allowing a direct numerical comparison between the two states. Results of exact diagonalization of finite disorder-free systems in the second Landau level show that, by increasing the short range component of the Coulomb interaction, the ground state undergoes a phase transition from the Pfaffian state to the compressed PH-Pfaffian state before further entering into a gapless state. The low energy gapped excited states result from the breakup of the abelian Laughlin type quasiparticle into two non-abelian quasiparticles.
\end{abstract}
 \maketitle

The understanding of the fractional quantum Hall effect (FQHE) at $\nu=5/2$ filling factor has undergone a roller coaster ride since it was discovered more than thirty years ago\cite{Willett}, and still remains a great challenge. The Pfaffian state\cite{MR} and its relevance to the $5/2$ FQHE was placed on a solid numerical ground and was once widely accepted as the most promising candidate by Morf's seminal work \cite{Morf} and other extensive studies followed \cite{Rezayi}\cite{Peterson}\cite{Wang}\cite{DasSarma}.  However, the dominant status of the Pfaffian state was shaken by the observation that it is not particle-hole (PH) symmetric, and together with its PH conjugate, termed anti-Pfaffian state, they form two degenerate but distinct states in the absence of Landau level mixing and disorder \cite{Levin} \cite{Lee}. Furthermore, recent numerical studies \cite{EHRezayi} show that when the Landau level mixing is properly taken into account, the degeneracy between the Pfaffian and anti-Pfaffian states is lifted in favor of the anti-Pfaffian state energetically, and the anti-Pfaffian state is therefore more likely to be the ground state of the $\nu=5/2$ FQHE. 

In the meantime, another topologically different state that is PH symmetric, termed PH-Pfaffian state, was suggested  by Son \cite{Son} in the context of Dirac composite fermion. Soon after,  two similar but different wave functions were proposed for the PH-Pfaffian state \cite{Zucker}\cite{Yang}. While these two proposed PH-Pfaffian wave functions occur at the total flux number $N_{\phi}= 2N-1$ and are shown to be highly PH symmetric in the spherical geometry, no consistent incompressible ground state is found to exist at such a $N_{\phi}$ and $N$ relationship even with a wide range of variations of Coulomb interactions\cite{Yang}. Questions are even raised if the PH-Pfaffian state represents a gapped, incompressible phase\cite{Mishmash}.
. 

While all the existing numerical results seem to converge to a consensus  that the anti-Pfaffian state is most likely the ground state of the $\nu=5/2$ FQHE, a recent breakthrough on the thermal Hall conductance measurement\cite{Banerjee} casts a great doubt on this consensus. The thermal Hall conductance is found to be $\kappa_{xy} = 5/2$ (in units of $\frac{\pi^2k^2_B}{3h}T$), which is incompatible with the edge structure of anti-Pfaffian, but rather consistent with the PH-Pfaffian topological order. This once again brings us back to the drawing board to search for a quantum Hall state that is supported by the numerical studies and the experiments. Nevertheless, the thermal Hall conductance measurement is rather encouraging and pointing to the existence of non-abelian quasiparticles, which is consistent with the Pfaffian, anti-Pfaffian, and PH-Pfaffian state.

In view of the discrepancy  between the numerical and experimental results, some alternative proposals are put forward to resolve the discrepancy. One compelling proposal is disorder induced mesoscopic puddles composed of Pfaffian and anti-Pfaffian states with effective edge structure that is consistent  with the thermal Hall conductance experiment\cite{Mross}\cite{CWang}\cite{Lian}. However, a recent more careful analysis on the energetics of forming such puddles ruled out the possibility\cite{Simon}.  There are other proposals such as incomplete thermal equilibration  on an anti-Pfaffian edge which is also considered to be unlikely\cite{Feldman}.

In this paper, we focus on a disorder free and no Landau level mixing system. In the spherical geometry, it is a well-established  fact that at the second Landau level, the relevant incompressible state occurs only at $N_{\phi}= 2N-3$ (or $N_{\phi}= 2N+1$ for equivalent  PH conjugate states). On the other hand, the PH-Pfaffian state can only be formed at $N_{\phi}= 2N-1$, however, no consistent incompressible ground states are found at such a magnetic flux number. We ask the following question: can we form an incompressible state at $N_{\phi}= 2N-3$ that is not PH symmetric yet maintains the PH-Pfaffian topological order and is energetically more favorable (or has larger overlap with the exact ground state) than the Pfaffian state, at least for a certain parameter range of the Coulomb interactions?

To that end we propose the following wave function:
\begin{equation}
\label{CPH-Pfaffian} {\Psi}_{CPH-Pf} = \int d^2{\xi}_1 d^2{\xi}_2 ({\xi}_1 -{\xi}_2)^N \prod\limits_{i=1}^N \prod\limits_{a=1}^2 (\frac{\partial}{{\partial}z_i} - {\xi}_a^*) {\Psi}_{PH-Pf}
\end{equation}
where
\begin{equation}
\label{PH-Pfaffian} {\Psi}_{PH-Pf} = {Pf} ( \frac{1}{ \frac{\partial}{{\partial}z_i}-\frac{\partial}{{\partial}z_j} } )  \prod\limits_{i<j}^N(\frac{\partial}{{\partial}z_i}-\frac{\partial}{{\partial}z_j} )  \prod\limits_{i<j}^N (z_i-z_j)^3
\end{equation}
is the PH-Pfaffian wave function \cite{Yang}, $z_j = x_j+iy_j$ is the complex coordinate of the $j_{th}$ electron, $N$ is the total number of electrons, and $Pf[A]$ is the Pfaffian of an antisymmetric matrix $A$. In Eq.(\ref{CPH-Pfaffian}),  $\prod\limits_{i=1}^N \prod\limits_{a=1}^2 (\frac{\partial}{{\partial}z_i} - {\xi}_a^*) $ creates two Laughlin type abelian quasiparticles located at ${\xi}_1$ and ${\xi}_2$ from the PH-Pfaffian state, and the two quasiparticles form a uniform state with the maximum avoidance from one another (or the maximum number, $N$, of zeros) in the form of $({\xi}_1 -{\xi}_2)^N$.  For the lack of a better term, we call the state a compressed PH-Pfaffian state, as it is formed from the PH-Pfaffian state by "compressing" it with two flux quanta removed. The compressed PH-Pfaffian state is not PH symmetric but resembles the PH-Pfaffian state, in particular, it possesses the PH-Pfaffian topological order. It is noted that the product of the Pfaffian and the Jastrow function of the derivatives in Eq.(\ref{PH-Pfaffian}) should be carried out first before applying to the function to its right, this way the derivatives appeared in the denominator in the Pfaffian will be cancelled out.

In Haldane's spherical geometry\cite{Haldane}, Eq.(\ref{PH-Pfaffian}) can be written as:
\begin{equation}
\label{SPH-Pfaffian} {\Psi}_{PH-Pf} = {Pf}(\frac{1}{  \frac{\partial}{{\partial}u_i}\frac{\partial}{{\partial}v_j}-\frac{\partial}{{\partial}u_j}\frac{\partial}{{\partial}v_i} })  \prod\limits_{i<j}^N(\frac{\partial}{{\partial}u_i}\frac{\partial}{{\partial}v_j}-\frac{\partial}{{\partial}u_j}\frac{\partial}{{\partial}v_i} ) {\Phi}_{3} 
\end{equation}
where ${\Phi}_{3} $ is the Laughlin wave function \cite{Laughlin}
\begin{equation}
\label{Laughlin_Sphere} {\Phi}_{3} =   \prod\limits_{i<j}^N (u_iv_j-u_jv_i)^{3} 
\end{equation}
and $(u, v)$ are the spinor variables describing electron coordinates. One can carry out the integration  over the quasiparticles coordinates and rewrite 
Eq.(\ref{CPH-Pfaffian}) in a much simpler form in the spherical geometry:
\begin{equation}
\label{SCPH-Pfaffian} {\Psi}_{CPH-Pf} = \sum\limits_{m=-\frac{N}{2}}^{\frac{N}{2}}  (-1)^m G_{m}G_{-m} {\Psi}_{PH-Pf}
\end{equation}
where 
\begin{eqnarray}
&&G_{m}=(-1)^{\frac{N}{2}-m}[\frac{N!}{(\frac{N}{2}+m)!(\frac{N}{2}-m)!}]^{-1/2}\cdot
 \nonumber \\
&&
\sum_{1{\leq}l_1<l_2<{\ldots}{\leq}l_{\frac{N}{2}+m}}\frac{\partial}{\partial
v_{l_1}}\frac{\partial}{\partial
v_{l_2}}\ldots\frac{\partial}{\partial v_{l_{\frac{N}{2}+m}}} \cdot\nonumber \\
&& \prod_{l({\neq} l_1, l_2, \ldots, l_{\frac{N}{2}+m})}
\frac{\partial}{\partial u_l}.
\end{eqnarray}
is a quasiparticle generation operator in angular momentum space \cite{YangHierarchy}, which
will generate a quasiparticle with angular momentum $(L,L_z)=(\frac{N}{2},m)$ when applied to ${\Psi}_{PH-Pf}$ which has angular momentum $L=0$. As a result, ${\Psi}_{CPH-Pf}$ is formed from the two quasiparticles  and has total angular momentum $L=0$, thus is rotationally invariant, a condition required for an incompressible state in the spherical geometry.

Since the total flux number $N_{\phi}$ corresponding to the PH-Pfaffian wave function is $N_{\phi} = 2N-1$, and the compressed PH-Pfaffian is formed from the PH-Pfaffian state by a removal of two flux quanta, the relationship between the flux number $N_{\phi}$ and the number of electrons $N$ is $N_{\phi} = 2N-3$. This is the same $N_{\phi}-N$ relationship\cite{Morf} for the Pfaffian state\cite{MR}
\begin{equation}
\label{Pfaffian} {\Psi}_{Pf} = {Pf} ( \frac{1}{z_i-z_j } )   \prod\limits_{i<j}^N (z_i-z_j)^2
\end{equation}
which has been studied extensively.

Since both wave functions ${\Psi}_{CPH-Pf}$ given by Eq.(\ref{CPH-Pfaffian}) and ${\Psi}_{Pf}$ given by Eq.(\ref{Pfaffian}) have the same $N_{\phi}-N$ relationship $N_{\phi} = 2N-3$, it allows for a direct numerical comparison between the two to determine which wave function, therefore which topological order and at what condition, represents the exact ground state.
In Fig.~\ref{fig:Overlap}, we calculated and plotted the overlap of the exact ground state of a finite system ($N_{\phi}, N$) = ($13, 8$) with the Pfaffian state and with the compressed PH-Pfaffian state respectively. The exact ground state is obtained in the second Landau level without disorder and Landau level mixing, with  the  ratios of $V_1/V_1^c$ ranging from $1$ to $1.5$, where $V_1^c$ is the Coulomb value of $V_1$ in 
the second Landau level. We see the ground state undergoes a phase transition from the Pfaffian state to the compressed PH-Pfaffian state  
as the short range component of the Coulomb interaction increases. The transition occurs at $V_1/V_1^c$ around $1.2$, as at this point the overlap of the exact ground state with the compressed PH-Pfaffian state exceeds that with the Pfaffian state. Therefore, if in the real system the Coulomb interaction  falls in a range such that the ground state is the compressed PH-Pfaffian state, the system will support the edge structure that is consistent with the thermal Hall conductance measurement.
 
\begin{figure}[tbhp]
\label{fig:Overlap}
%\vspace{0.2cm}
%\includegraphics[width=18cm,height=12cm]{Overlap_Pfaffian_QPH_Pfaffian_8e_QP}
\includegraphics[width=\columnwidth]{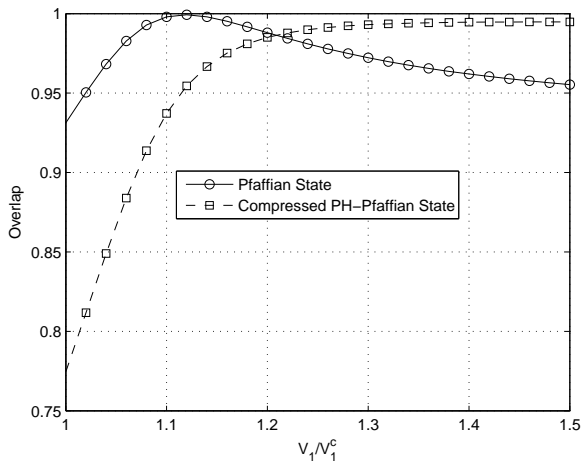}
\caption{\label{fig:Overlap} For $N=8$ and $N_{\phi} = 13$. Overlap of the exact ground state with the Pfaffian state (solid line) and the 
compressed PH-Pfaffian state (dashed line) as the function of the pseudopotential $V_1$
normalized by its Coulomb value $V_1^c$ in the second Landau level.}
\end{figure}

As there exists a PH conjugate state of the Pfaffian state, the anti-Pfaffian, there also exists a PH conjugate state of the compressed PH-Pfaffian state. Since the number of
electrons $N$ is related to the number  of holes $N_h$ of the PH conjugate state by $N+N_h = N_{\phi}+1$, the relationship between the flux number and the number of holes of the anti-Pfaffian or the PH conjugate state of the compressed PH-Pfaffian state is $N_{\phi} = 2N_h-1$. While Pfaffian and anti-Pfaffian are two topologically distinct states, we believe the compressed PH-Pfaffian state and its PH conjugate state have the same topological order. In the absence of PH symmetry breaking factors such as Landau level mixing, the same transition from the anti-Pfaffian state to the PH conjugate of the compressed PH-Pfaffian state will take place at the same short range interaction strength. 

One may also consider constructing   a "stretched" PH-Pfaffian state at $N_{\phi} = 2N+1$ from the PH-Pfaffian state by adding two flux quanta to create two abelian Laughlin type quasiholes
\begin{equation}
\label{CPH-Pfaffian1} \int d^2{\xi}_1 d^2{\xi}_2 ({\xi}_1^* -{\xi}_2^*)^N \prod\limits_{i=1}^N \prod\limits_{a=1}^2 (z_i - {\xi}_a) {\Psi}_{PH-Pf}
\end{equation}
Unfortunately, this stretched PH-Pfaffian state wave function has a rather poor overlap with the corresponding exact ground state even with an increased short range interaction. Fig.~\ref{fig:Overlap1} is the same as Fig.~\ref{fig:Overlap} except the overlap with the compressed PH-Pfaffian state is replaced by the overlap with the PH conjugate of the state described by Eq. (\ref{CPH-Pfaffian1}).
\begin{figure}[tbhp]
\label{fig:Overlap1}
%\vspace{0.2cm}
%\includegraphics[width=18cm,height=12cm]{Overlap_Pfaffian_QPH_Pfaffian_8e_QH}
\includegraphics[width=\columnwidth]{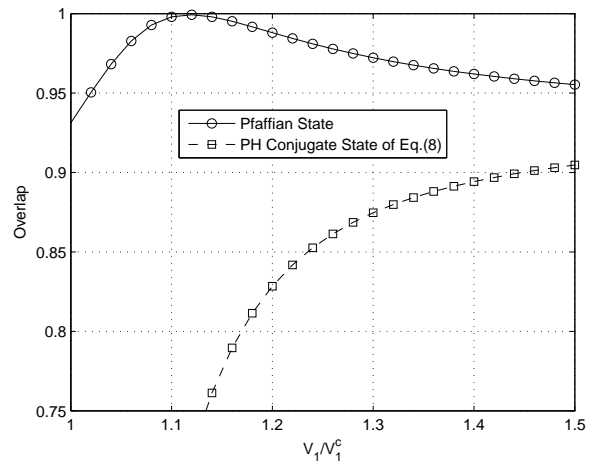}
\caption{\label{fig:Overlap1} Same as Fig.~\ref{fig:Overlap} except the overlap with compressed PH-Pfaffian state is replaced by the overlap with the PH conjugate of the state described by Eq. (\ref{CPH-Pfaffian1}).}
\end{figure}

Now we turn to the low energy excited states. A more general form of Eq.(\ref{SCPH-Pfaffian}) is:  
\begin{equation}
\label{SCPH-Pfaffian-L}\sum\limits_{m=-\frac{N}{2}}^{\frac{N}{2}}<L,0|\frac{N}{2},m;\frac{N}{2},-m>G_{m}G_{-m}{\Psi}_{PH-Pf}
\end{equation}
where $<L,m|L_1m_1;\L_2m_2>$ are the Clebsch-Gordon coefficients. Eq.(\ref{SCPH-Pfaffian-L}) not only describes the compressed PH-Pfaffian state, but also describe other states with even integer number of total angular momentum $L=2, 4, \cdot\cdot\cdot,N$.
Since the compressed PH-Pfaffian state provides a near unity overlap ($0.9931$) with the exact ground state for $N = 8$ when $V_1/V_1^c = 1.3$, one would expect that other states with non-zero angular momentum would also provide a good description for the low energy excited states. To the contrary, Eq.(\ref{SCPH-Pfaffian-L}) have rather poor overlaps with the corresponding exact low energy states: $0.5044$ at angular momentum $L = 2$,   $0.7723$ at $ L= 4$,  $0.4445$ at $L = 6$, and $0.8590$ at $L = 8$. This indicates that the low energy excited states are not formed from the two abelian Laughlin type quasiparticles. 

In the following, we will show that the low energy excited states are actually well described by the sub-Hilbert space spanned by four non-abelian quasiparticles  of the PH-Pfaffian state:
\begin{eqnarray}
\label{NonAbelianQP}
&&
{\Psi}({\xi}_1, {\xi}_2, {\xi}_3, {\xi}_4) = 
\nonumber \\
&&{Pf}(\frac{ \prod\limits_{a=1}^2 (\frac{\partial}{{\partial}z_i}- {\xi}_a^*) \prod\limits_{b=3}^{4} (\frac{\partial}{{\partial}z_j} - {\xi}_b^*) +(i {\leftrightarrow} j)}{ \frac{\partial}{{\partial}z_i}-\frac{\partial}{{\partial}z_j} })
\nonumber \\
&&\prod\limits_{i<j}^N(\frac{\partial}{{\partial}z_i}-\frac{\partial}{{\partial}z_j} )  \prod\limits_{i<j}^N (z_i-z_j)^3
\end{eqnarray}
The four non-abelian quasiparticles are located in ${\xi}_1$, ${\xi}_2$, ${\xi}_3$, and ${\xi}_4$, each carries $-\frac{1}{4}$ of the electron charge. In Eq.(\ref{NonAbelianQP}) the four non-abelian quasiparticles are arbitrarily divided into two groups, ${\xi}_1$ and ${\xi}_2$ are in one group, and ${\xi}_3$ and ${\xi}_4$ are in another. There are two other ways to divide the four quseiparticles into two groups, namely ${\xi}_1$ and ${\xi}_3$ in one group and ${\xi}_2$ and ${\xi}_4$ in another, or ${\xi}_1$ and ${\xi}_4$ in one group and ${\xi}_2$ and ${\xi}_3$ in another. One can symmetrize Eq.(\ref{NonAbelianQP}) with respect to ${\xi}_1$, ${\xi}_2$, ${\xi}_3$, and ${\xi}_4$:
\begin{equation}
\label{NonAbelianQP-Sum} {\Psi}({\xi}_1, {\xi}_2, {\xi}_3, {\xi}_4) + {\Psi}({\xi}_1, {\xi}_3, {\xi}_2, {\xi}_4) + {\Psi}({\xi}_1, {\xi}_4, {\xi}_2, {\xi}_3)
\end{equation}
expand Eq.(\ref{NonAbelianQP-Sum}) in terms of symmetrized polynomials of ${\xi}_1$, ${\xi}_2$, ${\xi}_3$, and ${\xi}_4$, and form independent basis functions in terms of the coordinates of the electrons. Since these basis functions are not orthogonal, we use Gram-Schmidt orthogonalization procedure to obtain an orthonomal set. We then diagonalize the Hamiltonian  in this subspace to obtain the energy spectrum and the wave functions. 

\begin{figure}[tbhp]
\label{fig:Overlap_4qp}
%\vspace{0.2cm}
%\includegraphics[width=18cm,height=12cm]{Overlap_4qp}
%\includegraphics[width=\columnwidth]{Overlap_4qp}
\includegraphics[width=\columnwidth]{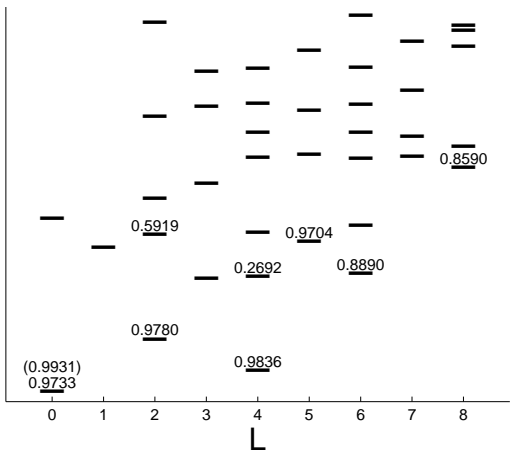}
\caption{\label{fig:Overlap_4qp} Energy spectrum of $(N_{\phi},N)=(13,8)$ finite system in the second Landau level with $V_1/V_1^c = 1.3$
in an arbitrary units versus angular momentum $L$. The numbers
on top of the energy bars are the
overlaps of the eigenstates of the Hamiltonian calculated in the subspace spanned by 
Eq.(\ref{NonAbelianQP-Sum})  with the corresponding exact states calculated in the full Hilbert space. }
\end{figure}

In Fig.~\ref{fig:Overlap_4qp}, we
plot the energy spectrum in an arbitrary units versus angular momentum $L$ for a
$(N_{\phi},N)=(13,8)$ finite system in the second Landau level with $V_1/V_1^c = 1.3$ by diagonalizing the Hamiltonian in the full Hilbert space. We also diagonalized the same Hamiltonian in the subspace spanned by Eq.(\ref{NonAbelianQP-Sum}). The overlaps of the eigenstates calculated in the subspace spanned by 
Eq.(\ref{NonAbelianQP-Sum})  and the corresponding exact states calculated in the full Hilbert space are also shown on top of the energy bars. For comparison, we also show the overlap ($0.9931$ within a parentheses) of the exact ground state with the compressed PH-Pfaffian state on top of the ground state energy bar at $L = 0$. The two different numbers on top of the ground state energy bar indicates that the compressed PH-Pfaffian state is not identical to the ground state calculated in the subspace spanned by  Eq.(\ref{NonAbelianQP-Sum}). In fact, their overlap is $0.9986$. The large overlaps of the two low energy excited states at $L = 2$ and $L = 4$ suggest that the low energy gapped excited states are formed from the breakup of the abelian Laughlin type quasiparticle into two non-abelian quasiparticles.

We have also diagonalized the Hamiltonian in a subspace spanned by the following two non-abelian and one abelian quasiparticles located in ${\xi}_1$, ${\xi}_2$, and ${\xi}_3$:
\begin{eqnarray}
\label{AbelianNonAbelianQP}
&&
{\Psi}({\xi}_1, {\xi}_2, {\xi}_3) = 
\nonumber \\
&&{Pf}(\frac{ (\frac{\partial}{{\partial}z_i}- {\xi}_1^*) (\frac{\partial}{{\partial}z_j} - {\xi}_2^*) }{ \frac{\partial}{{\partial}z_i}-\frac{\partial}{{\partial}z_j} })\prod\limits_{i<j}^N(\frac{\partial}{{\partial}z_i}-\frac{\partial}{{\partial}z_j} ) 
\nonumber \\
&&\prod\limits_{i=1}^N  (\frac{\partial}{{\partial}z_i} - {\xi}_3^*)  \prod\limits_{i<j}^N (z_i-z_j)^3
\end{eqnarray}
where the quasiparticles located at ${\xi}_1$ and ${\xi}_2$ are non-abelian and carry $\frac{1}{4}$ of the electron charge, and the quasiparticle located at ${\xi}_3$ is abelian and carries $\frac{1}{2}$ of the electron charge. One can symmetrize Eq.(\ref{AbelianNonAbelianQP}) with respect to ${\xi}_1$, ${\xi}_2$, and ${\xi}_3$:
\begin{equation}
\label{AbelianNonAbelianQP-Sum} {\Psi}({\xi}_1, {\xi}_2, {\xi}_3) + {\Psi}({\xi}_1, {\xi}_3, {\xi}_2) + {\Psi}({\xi}_2, {\xi}_3, {\xi}_1)
\end{equation}
and expand Eq.(\ref{AbelianNonAbelianQP-Sum}) in terms of symmetrized polynomials of ${\xi}_1$, ${\xi}_2$, ${\xi}_3$, and form independent basis functions in terms of the coordinates of the electrons. Since these basis functions are not orthogonal, we use Gram-Schmidt orthogonalization procedure to obtain an orthonomal set. We then diagonalize the Hamiltonian  in this subspace to obtain the energy spectrum and the eigenstates. For $N=8$ system, we found that the $L=0$ eigenstate in the subspace is identical to the compressed PH-Pfaffian state, other eigenstates in the subspace generally don't provide a good description for the low energy states. Some eigenstates in the subspace are even not the eigenstates of the angular momentum, indicating Eq.(\ref{AbelianNonAbelianQP-Sum}) does not form a complete sub-Hilbert space.

Before closing, we would like to point out that there exists an alternative PH-Pfaffian wave function \cite{Mross}:
\begin{equation}
\label{APH-Pfaffian} {\Psi}_{PH-Pf}' = {Pf} ( \frac{1}{z_i^*-z_j^* } )   \prod\limits_{i<j}^N (z_i-z_j)^2
\end{equation}
As a result, we can also write an alternative compressed PH-Pfaffian wave function in the following form:
\begin{equation}
\label{ACPH-Pfaffian} {\Psi}_{CPH-Pf}' = \int d^2{\xi}_1 d^2{\xi}_2 ({\xi}_1 -{\xi}_2)^N \prod\limits_{i=1}^N \prod\limits_{a=1}^2 (z_i^* - {\xi}_a^*) {\Psi}_{PH-Pf}'
\end{equation}
Eq.(\ref{CPH-Pfaffian}) and (\ref{PH-Pfaffian}) have advantages  over Eq.(\ref{ACPH-Pfaffian}) and (\ref{APH-Pfaffian}) in that they are defined in the lowest Landau level without a need of applying a projection operator, and therefore numerically easier  to handle. We don't expect the alternative wave functions will alter our results in any significant way.

In conclusion, we have proposed a compressed PH-Pfaffian state by "compressing" the PH-Pfaffian state with two flux quanta removed to create two abelian Laughlin type quasiparticles of the maximum avoidance from one another (or of the maximum number of zeros). The compressed PH-Pfaffian state is not particle-hole symmetric but possesses the PH-Pfaffian topological order. In the spherical geometry, results of exact diagonalization of finite disorder-free systems in the second Landau level show that, by increasing the short range component of the Coulomb interaction, the ground state undergoes a phase transition from the Pfaffian state to the compressed PH-Pfaffian state before further entering into a gapless state. The low energy gapped excited states are formed from the breakup of the abelian Laughlin type quasiparticle into two non-abelian quasiparticles.
More works are required on two fronts: First we need to see if the parameter range of the Coulomb interaction at which the system is in the compressed PH-Pfaffian state matches realistic conditions, in particular it is interesting to find out if the finite-thickness effects which alter all the pseudopotential components instead of just $V_1$\cite{Peterson} can provide such conditions. Secondly, we need to study larger size finite systems to verify if the compressed PH-Pfaffian state can survive  in the thermal dynamic limit.

\end{document}